\documentclass[reprint,amsmath,amssymb,aps,prd,]{revtex4-1}
\usepackage{graphicx}
\usepackage{epsfig,graphics,subfigure,psfrag,amsmath,amssymb}

\usepackage{dcolumn}
\usepackage{bm}
\usepackage{overpic}
\usepackage{xspace}

\usepackage{rotating}
\usepackage{color}
\usepackage{multirow}
\usepackage{colortbl}
\usepackage{epstopdf}
\usepackage{float}
\usepackage{lineno}
\definecolor{aa}{RGB}{0,0,139}
\usepackage[bookmarksnumbered=true,colorlinks,urlcolor=aa,linkcolor=blue,anchorcolor=aa,citecolor=aa]{hyperref}
\begin{document}
\normalsize
\parskip=5pt plus 1pt minus 1pt
\title{\boldmath Study of $\psi(3686)\rightarrow\Lambda\bar{\Lambda}\omega$}
\author{
\begin{small}
\begin{center}
M.~Ablikim$^{1}$, M.~N.~Achasov$^{10,b}$, P.~Adlarson$^{68}$, M.~Albrecht$^{4}$, R.~Aliberti$^{28}$, A.~Amoroso$^{67A,67C}$, M.~R.~An$^{32}$, Q.~An$^{64,50}$, X.~H.~Bai$^{58}$, Y.~Bai$^{49}$
% ...
\\
\vspace{0.2cm}
(BESIII Collaboration)\\
\vspace{0.2cm} {\it
$^{1}$ Institute of High Energy Physics, Beijing 100049, People's Republic of China\\
% ...
$^{4}$ Bochum Ruhr-University, D-44780 Bochum, Germany\\
% ...
$^{10}$ G.I. Budker Institute of Nuclear Physics SB RAS (BINP), Novosibirsk 630090, Russia\\
% ...
$^{28}$ Johannes Gutenberg University of Mainz, Johann-Joachim-Becher-Weg 45, D-55099 Mainz, Germany\\
% ...
$^{32}$ Liaoning Normal University, Dalian 116029, People's Republic of China\\
% ...
$^{49}$ Southeast University, Nanjing 211100, People's Republic of China\\
$^{50}$ State Key Laboratory of Particle Detection and Electronics, Beijing 100049, Hefei 230026, People's Republic of China\\
% ...
$^{58}$ University of Jinan, Jinan 250022, People's Republic of China\\
% ...
$^{64}$ University of Science and Technology of China, Hefei 230026, People's Republic of China\\
% ...
$^{67}$ University of Turin and INFN, (A)University of Turin, I-10125, Turin, Italy; (B)University of Eastern Piedmont, I-15121, Alessandria, Italy; (C)INFN, I-10125, Turin, Italy\\
$^{68}$ Uppsala University, Box 516, SE-75120 Uppsala, Sweden\\
% ...
$^{b}$ Also at the Novosibirsk State University, Novosibirsk, 630090, Russia\\
% ...
}
\end{center}
\vspace{0.4cm}
\end{small}
}

\setlength{\textfloatsep}{5pt}

\begin{abstract}

Based on a data sample of $(448.1\pm2.9)\times10^6$ $\psi(3686)$ events collected with the BESIII detector at the BEPCII collider, the branching fraction of $\psi(3686)\rightarrow\Lambda\bar{\Lambda}\omega$ is measured to be $\rm (3.30\pm0.34(stat.)\pm0.29(syst.))\times10^{-5}$ for the first time. In addition, the $\Lambda\omega$ (or $\bar{\Lambda}\omega$) invariant mass spectra is studied and the potential presence of excited $\Lambda$ states has been investigated.

\end{abstract}

\maketitle
\section{Introduction}

Quantum Chromodynamics (QCD), the theory which describes the strong
interaction, has been tested thoroughly at high energy. However, in
the medium energy region, theoretical calculations based on first
principles are still unreliable, since non-perturbative
contributions are significant and calculations have to rely on
models. Experimental measurements in this energy region are helpful to
validate models, constrain parameters, and inspire new
calculations. Charmonium states are on the boundary between the
perturbative and non-perturbative regimes; therefore, their decays,
especially hadronic decays, provide ideal inputs to the study of
QCD. The availability of very large data samples of vector charmonia,
such as $J/\psi$ and $\psi(3686)$, produced via electron-positron
annihilations, makes possible experimental studies of rare processes
and decay channels with complicated intermediate
structures~\cite{BESIII:2018gmc}.

Among these hadronic decays, scenarios of $\psi(3686)$ and $J/\psi$
decaying into baryon pairs have been understood in terms of $c\bar{c}$
annihilations into three gluons or into a virtual
photon~\cite{Zhu:2015bha}.  Three-body decays, namely
$\psi(3686)\rightarrow\Lambda\bar{\Lambda}P$, where $P$ represents a
meson such as $\pi^0$, $\eta$, or $\omega$, are of great interest
since intermediate states contribute
significantly~\cite{{BESIII:2012jve}}. Recent studies have focused
mainly on the final states $\Lambda\bar{\Lambda}\pi^0$ and
$\Lambda\bar{\Lambda}\eta$, and not so much on
$\Lambda\bar{\Lambda}\omega$, probably due to the fact that the final
states $\Lambda\bar{\Lambda}\pi^0$ and $\Lambda\bar{\Lambda}\eta$
allow to test the “12$\%$ rule”~\cite{BESIII:2018gmc}, but this is not
possible with $\Lambda\bar{\Lambda}\omega$, as it is heavily suppressed due to
the small phase space for the decay of $J/\psi$ to
$\Lambda\bar{\Lambda}\omega$. In addition, as the excitation spectra of
most hyperons are still not well understood~\cite{Sarantsev:2019xxm},
the process $\psi(3686)\rightarrow\Lambda\bar{\Lambda}\omega$ provides
a good opportunity to search for potential $\Lambda$ excitations. Multichannel analysis of the quasi-two-body excited hyperon states have been performed in the past: a pole and Breit-Wigner resonance parameters, assigned to $\Lambda(2070)\rightarrow\Lambda\omega$ are given as ${\rm M}=2044\pm20~{\rm MeV}/c^2$, ${\rm \Gamma}=360\pm45~{\rm MeV}$ and ${\rm M}=2070\pm24~{\rm MeV}/c^2$, ${\rm \Gamma}=370\pm50~{\rm MeV}$, respectively~\cite{Sarantsev:2019xxm}.

Using the large sample of $(448.1\pm2.9)\times10^6$ $\psi(3686)$
events collected with the BESIII detector, we present the first
observation of $\psi(3686)\rightarrow\Lambda\bar{\Lambda}\omega$ and
the corresponding upper limit of the branching fraction of
${\cal~B}(\psi(3686)\rightarrow\Lambda\bar{\Lambda}^{\ast}+c.c~\rightarrow\Lambda\bar{\Lambda}\omega)$.

\section{Detector and Data Samples}

The BESIII detector~\cite{Ablikim:2009aa} records symmetric $e^+e^-$ collisions  provided by the BEPCII storage ring~\cite{Yu:IPAC2016-TUYA01}, which operates with a peak luminosity of $1\times10^{33}$~cm$^{-2}$s$^{-1}$ in the center-of-mass~(CM) energy range from 2.0 to 4.9~GeV. BESIII has collected large data samples in this energy region~\cite{BESIII:2020nme}. The cylindrical core of the BESIII detector covers 93\% of the full solid angle and consists of a helium-based multilayer drift chamber~(MDC), a plastic scintillator time-of-flight system~(TOF), and a CsI(Tl) electromagnetic calorimeter~(EMC), which are all enclosed in a superconducting solenoidal magnet providing a 1.0~T magnetic field. The solenoid is supported by an octagonal flux-return yoke with resistive plate counter muon identification modules interleaved with steel.
The charged-particle momentum resolution at $1~{\rm GeV}/c$ is
$0.5\%$, and the specific energy loss ($dE/dx$) resolution is $6\%$ for electrons from Bhabha scattering. The EMC measures photon energies with a resolution of $2.5\%$ ($5\%$) at $1$~GeV in the barrel (end cap) region. The time resolution in the TOF barrel region is 68~ps, while that in the end cap region is 110~ps.

Simulated data samples produced with {\sc
  geant4}-based~\cite{GEANT4:2002zbu} Monte Carlo (MC) software, which
includes the geometric description of the BESIII detector and the
detector response, are used to optimize the selection criteria,
determine detection efficiencies and to estimate backgrounds. The
simulation models the beam energy spread and initial state radiation
(ISR) in the $e^+e^-$ annihilations with the generator {\sc
  kkmc}~\cite{Jadach:2000ir,Jadach:1999vf}. The inclusive MC
simulation includes the production of the $\psi(3686)$ resonance, the
ISR production of the $J/\psi$, and the continuum processes
incorporated in {\sc kkmc}. The known decay modes are modeled with
{\sc evtgen}~\cite{Lange:2001uf,Ping:2008zz} using branching fractions
taken from the Particle Data Group~\cite{Zyla:2020zbs}, and the
remaining unknown charmonium decays are modeled with {\sc
  lundcharm}~\cite{Chen:2000tv,Yang:2014vra}. Final state
radiation~(FSR) from charged final state particles is incorporated
using {\sc photos}~\cite{Richter-Was:1992hxq}.  An exclusive phase
space MC is used to simulate the
$\psi(3686)\rightarrow\Lambda\bar{\Lambda}\omega$ reaction. The data
sets collected at the CM energies of 3.650 and 3.773 GeV, with an
integrated luminosity of $(44.49\pm0.02\pm0.44)~{\rm pb^{-1}}$ and
$(2917\pm29)~{\rm pb^{-1}}$~\cite{Ablikim:2013ntc}, respectively, are
used to estimate the contamination from the continuum processes of
$e^{+}e^{-}\rightarrow\Lambda\bar{\Lambda}\omega$.

\section{Event Selection and Background Analysis}

The process $\psi(3686)\rightarrow\Lambda\bar{\Lambda}\omega$ is
reconstructed with $\Lambda\rightarrow~p\pi^{-}$,
$\bar{\Lambda}\rightarrow~\bar{p}\pi^{+}$,
$\omega\rightarrow~\pi^{+}\pi^{-}\pi^{0}$ and
$\pi^{0}\rightarrow~\gamma\gamma$. The signal events are required to
have at least six charged tracks and at least two photon
candidates. Charged tracks detected in the MDC are required to be
within a polar angle ($\theta$) range of $|\!\cos\theta|<0.93$, where
$\theta$ is defined with respect to the $z$-axis, which is the
symmetry axis of the MDC.  A secondary vertex fit is performed for
each pair of oppositely charged tracks, and the combinations with
$\chi^{2}$ less than 50 are saved as $\Lambda$ or $\bar{\Lambda}$
candidates, without further requirements on the decay length. At least
one $\Lambda\bar{\Lambda}$ pair is required in each signal event.
Particle identification (PID) based on TOF and $dE/dx$ information is
applied to the pairs of charged tracks belonging to the $\Lambda$ (
$\bar{\Lambda}$) candidate. The track in the pair with the largest
probability calculated under the proton (anti-proton) assumption is
assumed to be a proton (anti-proton), and the other track is assumed
to be a pion. If more than one $\Lambda \bar{\Lambda}$ pair survive,
the one with the minimum value of
($(M_{p\pi^{-}}-m_{\Lambda})^2+(M_{\bar{p}\pi^{+}}-m_{\bar{\Lambda}})^2$)
is kept for further analysis, where
$M_{p\pi^{-}}$~($M_{\bar{p}\pi^{+}}$) is the invariant mass of
$p\pi^{-}~(\bar{p}\pi^{+})$ and $m_{\Lambda}$~($m_{\bar{\Lambda}}$) is
the known mass of $\Lambda$~($\bar{\Lambda}$)~\cite{Zyla:2020zbs}.
The charged tracks which are not used in the $\Lambda\bar{\Lambda}$
pairs are assumed to originate from $\omega$ decays. For these tracks,
the distance of closest approach to the $e^{+}e^{-}$ interaction point
(IP) must be less than 10\,cm in the $z$ direction, and less than
1\,cm in the plane perpendicular to the $z$ direction.  They are
identified as $\pi$ if the pion PID likelihood is the largest among
the three assumptions of $\pi$, $K$ and $p$. Signal events are
selected when at least one pair of $\pi^{+}\pi^{-}$ is identified.

Photon candidates are identified using showers in the EMC. The
deposited energy of each shower must be greater than 25~MeV in the
barrel region ($|\!\cos\theta|< 0.80$) and greater than 50~MeV in the
end cap region ($0.86 <|\!\cos\theta|< 0.92$). To exclude showers
originating from charged particles, the angle between the position of
the shower in the EMC and the closest point of the extrapolated
trajectory of each charged track must be greater than 10 degrees;
because of  the large number of secondary photons produced in anti-proton
annihilations~\cite{BESIII:2017qwj}, the angle between the position of
the shower and the extrapolated trajectory point of anti-proton tracks
must be greater than 30 degrees. To suppress electronic noise and
showers unrelated to the event, the difference between the EMC time
and the event start time is required to be within~(0, 700)\,ns. The
$\pi^{0}$ candidates are reconstructed through $\pi^0\to \gamma\gamma$
decays, and a kinematic fit that constrains the invariant mass to the
$\pi^0$ known mass is performed~\cite{Zyla:2020zbs}. The $\chi^2$ of
the kinematic fit is required to be less than 25.

\begin{figure}[t]
\vspace{-0.5 cm} 
\setlength{\abovecaptionskip}{0 cm}  
\centering
  \hbox{
  \begin{overpic}[width=0.23\textwidth]{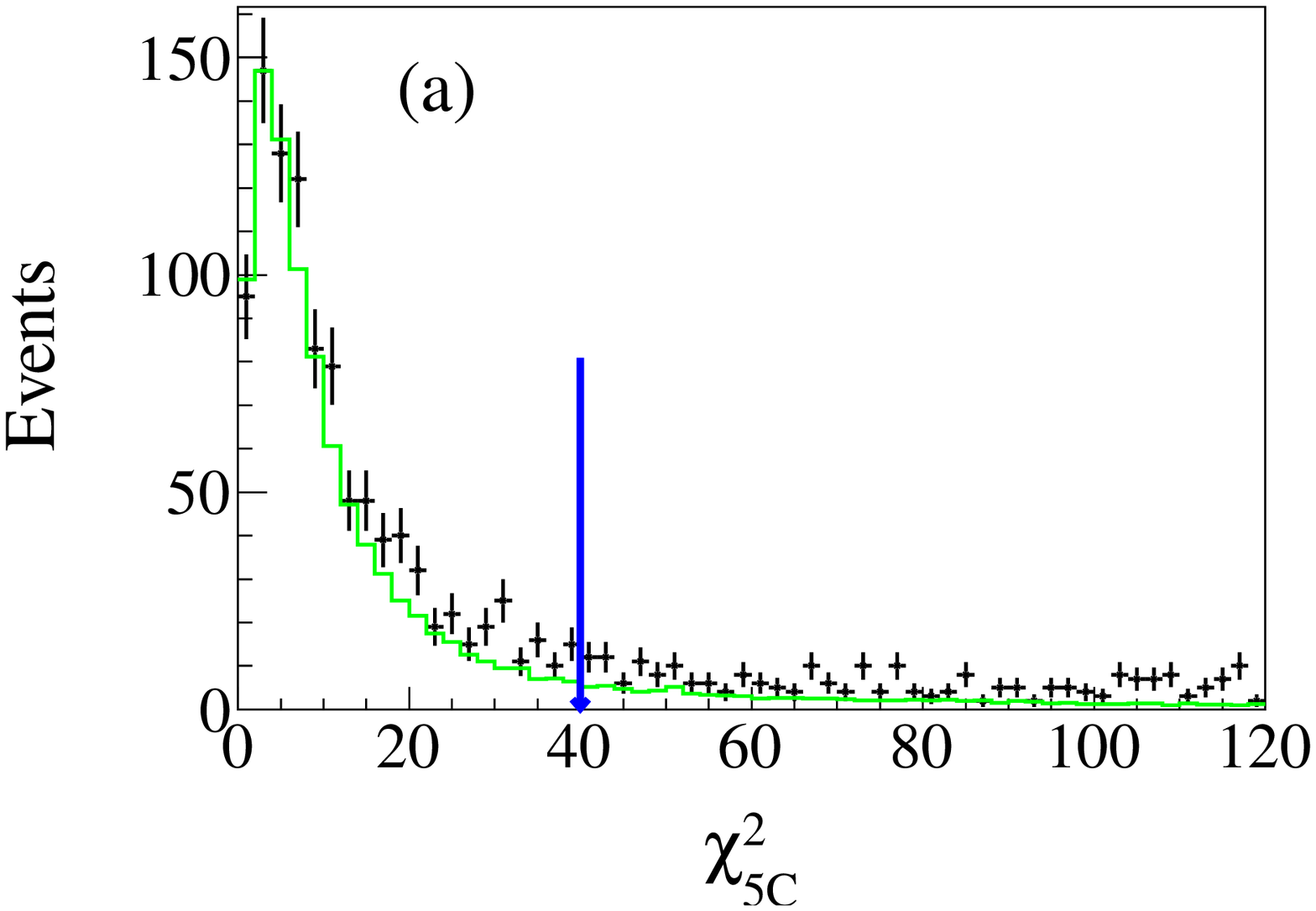}
  \put(50,1){}
  \end{overpic}
  \begin{overpic}[width=0.23\textwidth]{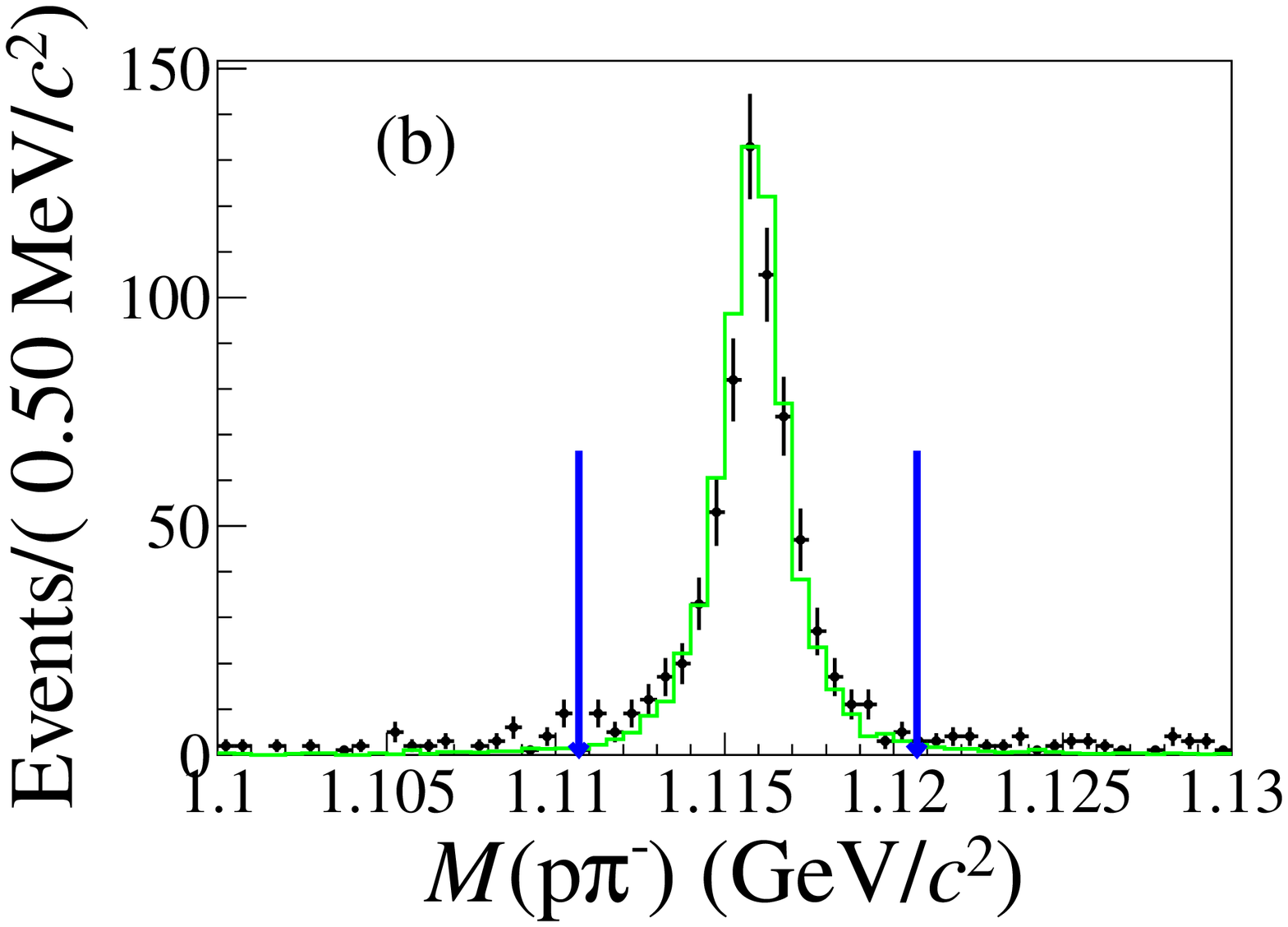}
  \put(50,1){}
  \end{overpic}
  }
  \centering
  \hbox{
  \begin{overpic}[width=0.23\textwidth]{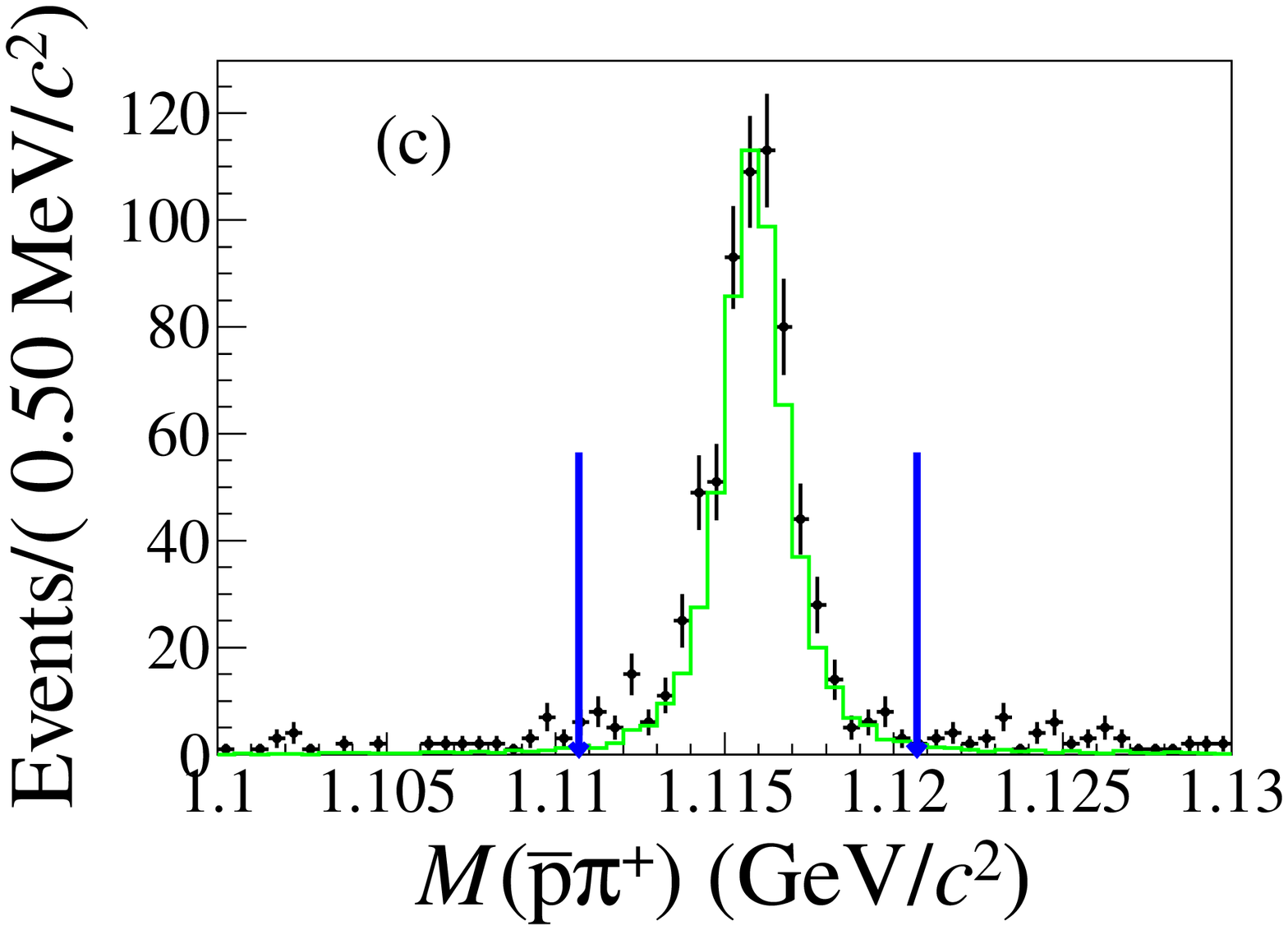}
  \put(50,0){}
  \end{overpic}
  \begin{overpic}[width=0.23\textwidth]{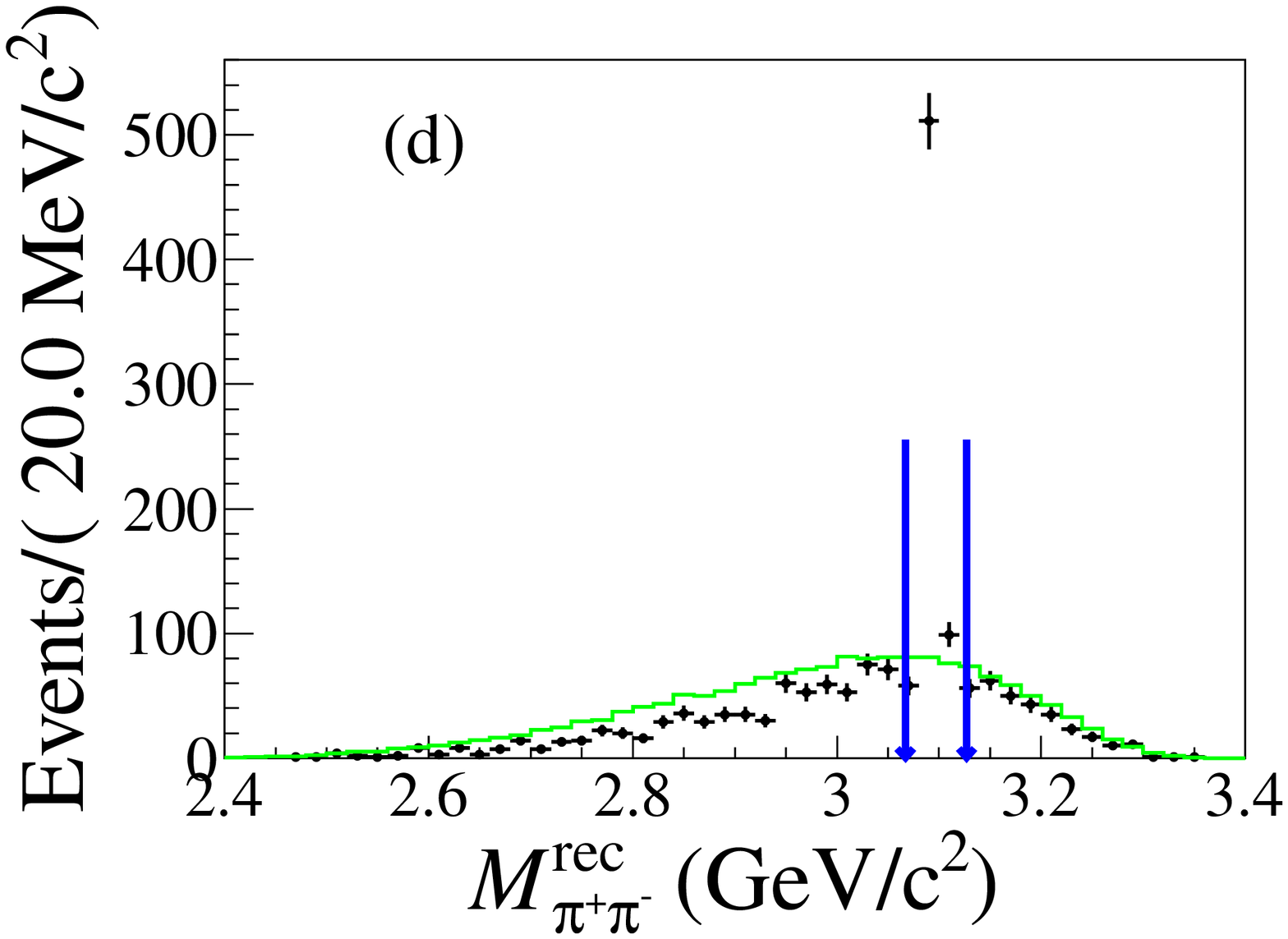}
  \put(50,0){}
  \end{overpic}
  }
 \caption{(a) The $\chi^{2}$ distribution of the 5C fit, (b) the
   invariant mass distribution of $p\pi^{-}$ and (c) $\bar{p}\pi^{+}$,
   and (d) the recoil mass of the $\pi^{+}\pi^{-}$ system from the
   decaying $\omega$. Dots with error bars are
   $\psi(3686)$ data, and the green histograms show the phase space MC
   simulation of
   $\psi(3686)\rightarrow\Lambda\bar{\Lambda}\omega$. The blue arrows
   show the applied selections. The simulation is normalized to the
   maximum bin value of the data in (a),(b) and (c), and to
   the data integral in (d). }
 \label{cutfigures}
\end{figure}

To improve the mass resolution, a five-constraint (5C) kinematic fit
for the $\psi(3686)\rightarrow
\Lambda\bar{\Lambda}\pi^{+}\pi^{-}\gamma\gamma$ hypothesis is
performed, where in addition to energy-momentum conservation, the
invariant mass of the two photons is constrained to the known mass of
$\pi^{0}$.  Figure~$\ref{cutfigures}$(a) shows the $\chi^{2}$ of the 5C
fit, which is required to be less than 40. For events with more than
one combination satisfying this requirement, the combination with the
smallest $\chi^{2}$ is kept.

For the surviving events, $\Lambda(\bar{\Lambda})$ candidates are
selected by requiring
$|M_{p\pi^{-}(\bar{p}\pi^{+})}-m_{\Lambda(\bar{\Lambda})}|<5~{\rm
  MeV}/c^{2}$, as shown in Fig.~$\ref{cutfigures}$(b) and
Fig.~$\ref{cutfigures}$(c). Backgrounds from
$\psi(3686)\rightarrow~\pi^{+}\pi^{-}J/\psi$ decays are rejected by
requiring $|M^\textrm{rec}_{\pi^{+}\pi^{-}}-m_{J/\psi}|>30~{\rm
  MeV}/c^{2}$, where $M^\textrm{rec}_{\pi^{+}\pi^{-}}$ is the recoil
mass of the $\pi^{+}\pi^{-}$ system decaying from $\omega$ and
$m_{J/\psi}$ is the known mass of $J/\psi$~\cite{Zyla:2020zbs}, as
shown in Fig.~$\ref{cutfigures}$(d).

To study the background contributions, the same selection criteria are
applied to the $\psi(3686)$ inclusive MC simulation. A topological
analysis of the events surviving to the selections has been performed
with TopoAna~\cite{Zhou:2020ksj}, and the number of
background events with $\omega$ in the final state is negligible.
Therefore, the background contributions are estimated using the
$\omega$ sideband events from data, where the sideband regions are
defined as $0.693~{\rm GeV}/c^2 < M_{\pi^+\pi^-\pi^0} < 0.723~{\rm
  GeV}/c^2$ and $0.843~{\rm GeV}/c^2 < M_{\pi^+\pi^-\pi^0} <
0.873~{\rm GeV}/c^2$, and the signal region as $0.753~{\rm GeV}/c^2 <
M_{\pi^+\pi^-\pi^0} < 0.813~{\rm GeV}/c^2$.  To investigate the amount
of possible background from continuum processes, the same selection
criteria are applied to data samples of $(2917\pm29)~{\rm pb^{-1}}$
collected at $\sqrt{s}=3.773~{\rm GeV}$ and
$(44.49\pm0.02\pm0.44)~{\rm pb^{-1}}$ collected at
$\sqrt{s}=3.650~{\rm GeV}$~\cite{Ablikim:2013ntc}, and no event
survives after applying all the selection criteria~\cite{BESIII:2019efv}.
Hence, background from the continuum contribution of the
electromagnetic process
$e^{+}e^{-}\rightarrow\Lambda\bar{\Lambda}\omega$ is negligible.

\section{\boldmath Branching Fraction Measurement of $\psi(3686)\rightarrow\Lambda\bar{\Lambda}\omega$}

\begin{figure}[H]
\setlength{\abovecaptionskip}{0 cm}   
\setlength{\belowcaptionskip}{0 cm}   
	\centering
	\includegraphics[width=0.44\textwidth]{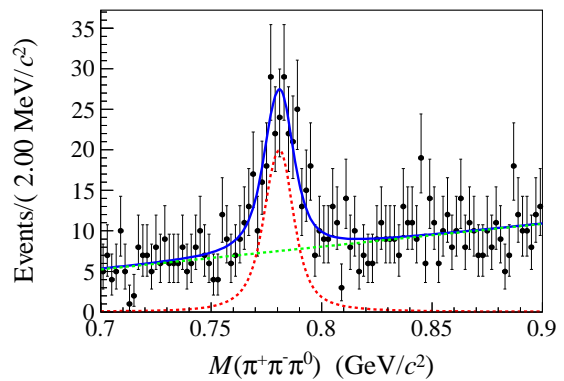}\\
	\caption{ Invariant mass spectrum of the $\pi^{+}\pi^{-}\pi^{0}$
          system. Dots with error bars are data, and the blue solid
          curve shows the fit result. The red dashed curve is the
          $\omega$ signal shape, and the green dashed curve is the
          background.}
	\label{fitomega}
\end{figure}

An unbinned maximum likelihood fit is performed to the invariant mass
distribution of $\pi^{+}\pi^{-}\pi^{0}$, as shown in
Fig.~$\ref{fitomega}$, to determine the number of $\omega$ signal
events. The $\omega$ signal shape is taken from signal MC simulation,
and the background is described by a first order Chebyshev polynomial.
The branching fraction (BF) of $\psi(3686)\rightarrow\Lambda\bar{\Lambda}\omega$ is
calculated according to:
\begin{equation}
{\cal B}(\psi(3686)\rightarrow\Lambda\bar{\Lambda}\omega)=\frac{{\rm N}_{\textrm{obs}}}{{\rm N}_{\psi(3686)} \cdot {\cal B}\cdot \varepsilon },
\label{eq1}
\end{equation}
where ${\rm N}_{\textrm{obs}}=207\pm21$ is the number of $\omega$
signal events obtained from the fit, $\varepsilon=(3.89\pm0.02)\%$ is
the detection efficiency, estimated from
$\psi(3686)\rightarrow\Lambda\bar{\Lambda}\omega$ MC simulation, and
${\cal B}$ is the product of branching ratios of ${\cal
  B}(\Lambda\rightarrow p\pi^{-}) \cdot {\cal
  B}(\bar{\Lambda}\rightarrow \bar{p}\pi^{+}) \cdot {\cal
  B}(\omega\rightarrow \pi^{+}\pi^{-}\pi^{0})\cdot {\cal
  B}(\pi^{0}\rightarrow \gamma\gamma)$~\cite{Zyla:2020zbs}. ${\cal
  B}(\psi(3686)\rightarrow\Lambda\bar{\Lambda}\omega)$ is measured to
be $\rm (3.30\pm0.34 (stat.))\times10^{-5}$, where the uncertainty is
statistical only.

\section{Study of Intermediate States}
\begin{figure}[t]
\setlength{\abovecaptionskip}{0 cm} 
\setlength{\belowcaptionskip}{0 cm} 
	\centering
	\includegraphics[width=0.43\textwidth]{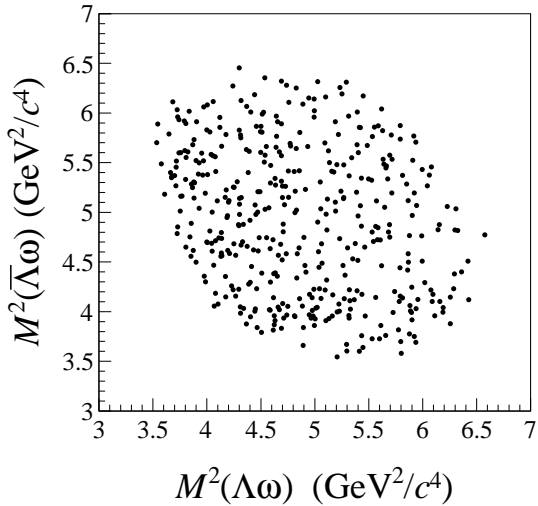}
	\caption{Dalitz plot of $M^{2}(\bar{\Lambda}\omega)$ versus $M^{2}(\Lambda\omega)$ from data.}
	\label{dalitz}
\end{figure}

In order to study possible intermediate states in
$\psi(3686)\rightarrow\Lambda\bar{\Lambda}\omega$, an additional
requirement of $|M_{\pi^{+}\pi^{-}\pi^{0}}-m_{\omega}|< 30~{\rm
  MeV}/c^{2}$ is applied, where $m_{\omega}$ is the mass of the
$\omega$ meson~\cite{Zyla:2020zbs}. The Dalitz plot of
$M^{2}(\bar{\Lambda}\omega)$ versus $M^{2}(\Lambda\omega)$ is
shown in Fig.~\ref{dalitz}. A two-dimensional unbinned maximum likelihood fit is performed to the Dalitz plot to investigate the potential exicted Lambda state $\Lambda^{*}$, which could decay as $\Lambda^{\ast}\rightarrow\Lambda\omega(\bar{\Lambda^{\ast}}\rightarrow\bar{\Lambda}\omega)$. In the fit, the signal of $\Lambda^{*}$ is modeled
with an S-wave Breit-Wigner function in two
dimensions~\cite{BESIII:2017vtc}, namely:
\begin{eqnarray}
  \epsilon(x,y)\cdot(\frac{p{\cdot}q}{(M^{2}_{R}
    -x)^{2}+M^{2}_{R}\cdot\Gamma^{2}} \nonumber \\
 + \frac{p{\cdot}q}{(M^{2}_{R} -y)^{2}+M^{2}_{R}\cdot\Gamma^{2}})\otimes\sigma(x,y)
  \label{eq2}
\end{eqnarray}
where $x$ and $y$ correspond to $M^{2}(\Lambda\omega)$ and $M^{2}(\bar{\Lambda}\omega)$, respectively; $M_R$ and $\Gamma$ are the mass and width of $\Lambda^{\ast}/\bar{\Lambda}^{\ast}$; $p$ is the momentum of $\Lambda^{\ast}/\bar{\Lambda}^{\ast}$ in the c.m. frame; $q$ is the momentum of $\omega$ in the rest frame of $\Lambda^{\ast}/\bar{\Lambda}^{\ast}$; $\sigma(x,y)$, is the two dimensional Gaussian resolution function, obtained from a zero width $\Lambda^{\ast}$ and $\bar{\Lambda}^{\ast}$ signal MC simulation, and $\epsilon(x,y)$ is the two dimensional detection efficiency function of $x$ and $y$, which is obtained from $\psi(3686)\rightarrow\Lambda\bar{\Lambda}\omega$ exclusive MC simulation. Due to the low statistics, possible interference effects are not considered in the analysis. The probability density function of the process $\psi(3686)\rightarrow\Lambda\bar{\Lambda}\omega$ without intermediate states is taken from MC simulation. The non$-\omega$ background is described by the normalized $\omega$ sidebands.

\begin{figure}[t]
\setlength{\abovecaptionskip}{0 cm}
\setlength{\belowcaptionskip}{0 cm}
\centering
  \hbox{
  \begin{overpic}[width=0.44\textwidth]{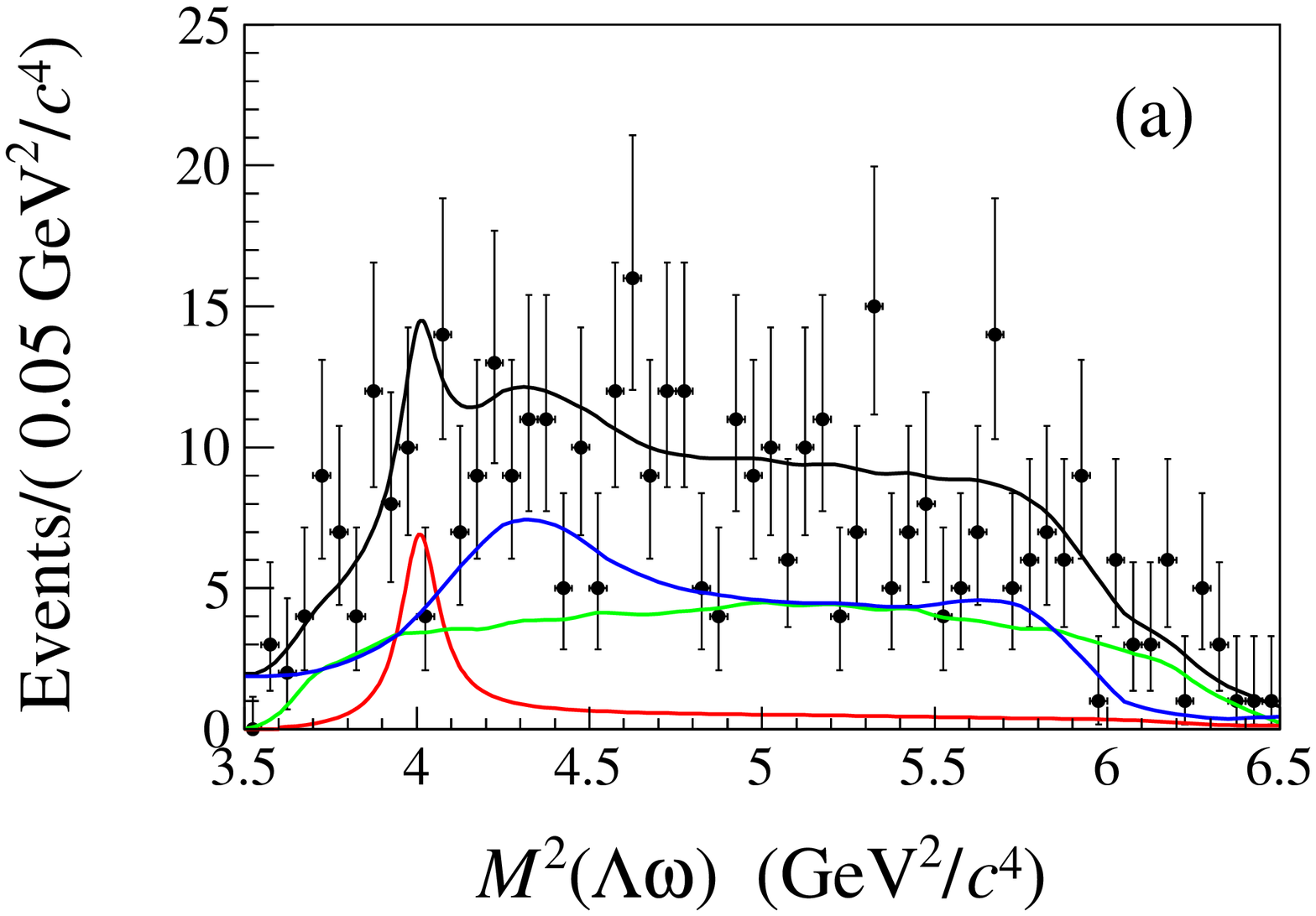}
  \put(50,1){}
  \end{overpic}
  }
 \centering
  \hbox{
   \begin{overpic}[width=0.44\textwidth]{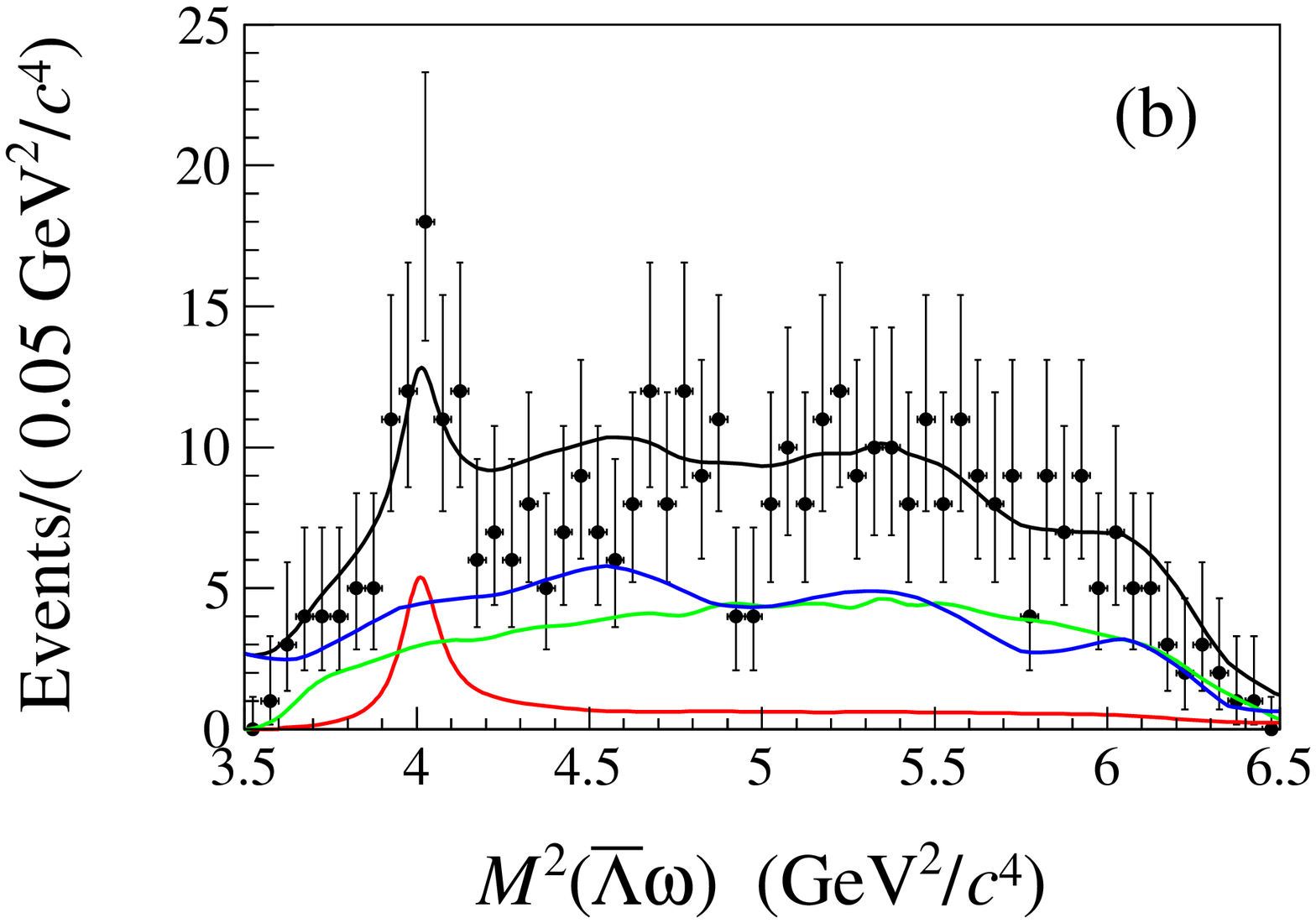}
  \put(50,0){}
  \end{overpic}
  }
 \caption{The distributions of (a) $M^{2}(\Lambda\omega)$ (b) and
   $M^{2}(\bar{\Lambda}\omega)$. Dots with error bars are
   data, the black solid curves are projections from the fits, the red
   solid curves show the shape of
   $\Lambda^{\ast}/\bar{\Lambda}^{\ast}$ resonances, the blue solid
   curves show the background described by $\omega$ sidebands and the
   green solid curves show the shapes from the non-resonant decay
   $\psi(3686)\rightarrow\Lambda\bar{\Lambda}\omega$. The $\Lambda\omega$ and $\bar{\Lambda}\omega$ invariant masses are fitted simultaneously giving the same yield for the $\Lambda\omega$ and $\bar{\Lambda}\omega$ excited state. The differences between the $\Lambda\omega$ and $\bar{\Lambda}\omega$ invariant masses in the fitted curves in (a) and (b) for the signal and the background are due to different detection efficiencies of $\Lambda\omega$ and $\bar{\Lambda}\omega$.}
	\label{fitlambdastar}
\end{figure}

The fit projections of $M^{2}(\Lambda\omega)$ and
$M^{2}(\bar{\Lambda}\omega)$ are shown in
Fig.~\ref{fitlambdastar}. From the fit, $51\pm16$ signals are obtained, with statistical signficance of  3.1$\sigma$, which is evaluated by comparing the likelihood values with and without the $\Lambda^{\ast}$ resonance included in the fit. To take into account the additive systematic uncertainties related to the fits, alternative fits with different $\omega$ sideband background levels and shapes are performed, and the minimum significance among these cases is 3.0$\sigma$. The fitted
$\Lambda^{\ast}/\bar{\Lambda}^{\ast}$ mass and width are
$M_{\Lambda^{\ast}/\bar{\Lambda}^{\ast}}$=$2.001\pm0.007~{\rm
  GeV}/c^2$, $\Gamma_{\Lambda^{\ast}/\bar{\Lambda}^{\ast}}$=$0.036\pm0.014~{\rm
  GeV}/c^2$.
If compared with the excited states with similar mass listed in the Review of Particle Properties~\cite{Zyla:2020zbs}, like $\Lambda(2000)$, $\Lambda(2050)$ and $\Lambda(2070)$, the fitted width of $36~{\rm MeV}/c^2$ is below the reported value of these states, which is between 100 and 400 ${\rm MeV}/c^2$. It is evident that higher statistics and a detailed partial wave analysis are needed in order to decide whether this structure is present and if its properties are consistent with one of these states or not. 

 The BF of $\psi(3686)\rightarrow\Lambda\bar{\Lambda}^{\ast}+c.c~\rightarrow\Lambda\bar{\Lambda}\omega$ according to 
\begin{equation}
\begin{aligned}
{\cal B}(\psi(3686)\rightarrow\Lambda\bar{\Lambda}^{\ast}+c.c~\rightarrow\Lambda\bar{\Lambda}\omega)=
\frac{{\rm N}_{\textrm{sig}}}{{\rm N}_{\psi(3686)} \cdot {\cal B}\cdot {\cal \epsilon}} \\
=(8.91\pm 2.82(stat.)\pm0.73(sys.))\times10^{-6},
\label{eq_br_lamstar}
\end{aligned}
\end{equation}
where ${\rm N}_{\textrm{sig}}=51\pm16$ is the number of $\Lambda^{*}$ signal events obtained from the fit, $\varepsilon=(3.59\pm0.02)\%$ is the
detection efficiency, estimated from
$\psi(3686)\rightarrow\Lambda\bar{\Lambda}^{\ast}+c.c~\rightarrow\Lambda\bar{\Lambda}\omega$ MC simulation and ${\cal B}$ is the product of branching ratios as Eq.~\ref{eq_br_lamstar}. Since only a few signal events are observed and the significance is only 3$\sigma$, to be conservative, the branching fraction upper limit of
$\psi(3686)\rightarrow\Lambda\bar{\Lambda}^{\ast}+c.c~\rightarrow\Lambda\bar{\Lambda}\omega$ is measured. The upper limit on the number of signal events,
${\rm N}_{\textrm{sig}}^{\textrm{up}}$, is determined at the 90$\%$
confidence level (CL) by solving~\cite{BESIII:2018gvg}:
\begin{equation}
\int^{{\rm N}_{\textrm{sig}}^{\textrm{up}}}_{0}{\cal L}(\mu)~d\mu/\int^{+\infty}_{0}{\cal L}(\mu)~d\mu=0.9,
\label{eq3}
\end{equation}
where $\mu$ is the number of fitted signal events, and ${\cal L}(\mu)$
is the likelihood function obtained from the fit to data. To determine the upper limit on the $\Lambda^{\ast}$ resonance signal events, a series of unbinned maximum-likelihood fits are performed to the Dalitz plot of $M^{2}(\bar{\Lambda}\omega)$ versus $M^{2}(\Lambda\omega)$ with a varying number of expected
  $\Lambda^{\ast}$ resonance signal events. To take into account the additive systematic uncertainties related to the fits, alternative fits with different $\omega$ sideband background levels and shapes are performed, and the maximum upper limit among these cases is
  determined. To account for the multiplicative systematic uncertainties, the likelihood distribution is convolved with a Gaussian function $G(x;0,\sigma)$ with a standard deviation of $\sigma=x\times\Delta$~\cite{BESIII:2018gvg}:
\begin{equation}
{\rm \cal  L^{'}(\mu)}=\int^{+\infty}_{0}{\cal L}(x)\times G(\mu-x;0,\sigma)\, dx
\label{eq4}
\end{equation}
where $\mu$ is the expected number of signal events, ${\rm \cal
  L^{'}(\mu)}$ indicates the expected likelihood distribution, and
$\Delta$ refers to the total relative systematic uncertainty listed
in Table~\ref{sumsysuncertainties}.

\begin{figure}[H]
\setlength{\abovecaptionskip}{0.cm}
\setlength{\belowcaptionskip}{-0.cm}
\centering
\vspace{-0.3cm}
  \hbox{
  \begin{overpic}[width=0.46\textwidth]{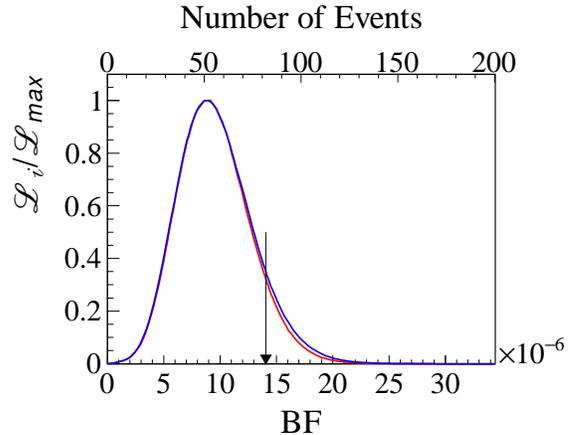}
  \put(50,1){}
  \end{overpic}
  }
 \caption{Distribution of likelihood versus number of signal events (and BF) for data. The results obtained with and without incorporating the systematic uncertainties are shown in blue solid and red solid curves, respectively. The black arrow shows the results corresponding to the $90\%$ CL.}
 \label{uplimits}
\end{figure}

The branching fraction upper limit of
$\psi(3686)\rightarrow\Lambda\bar{\Lambda}^{\ast}+c.c~\rightarrow\Lambda\bar{\Lambda}\omega$
is calculated as follows:
\begin{equation}
\begin{aligned}
{\cal B}(\psi(3686)\rightarrow\Lambda\bar{\Lambda}^{\ast}+c.c~\rightarrow\Lambda\bar{\Lambda}\omega)<
	\frac{{\rm N}_{\textrm{sig}}^{\textrm{up}}}{{\rm N}_{\psi(3686)} \cdot {\cal B}\cdot {\cal \epsilon}},
 \label{eq5}
\end{aligned}
\end{equation}
where ${\rm N}_{\textrm{sig}}^{\textrm{up}}$ is the upper limit of the
number of signal events, $\varepsilon=(3.59\pm0.02)\%$ is the
detection efficiency, estimated from
$\psi(3686)\rightarrow\Lambda\bar{\Lambda}^{\ast}+c.c~\rightarrow\Lambda\bar{\Lambda}\omega$ MC simulation, and
${\cal B}$ is the product of branching ratios of ${\cal
  B}(\Lambda\rightarrow p\pi^{-}) \cdot {\cal
  B}(\bar{\Lambda}\rightarrow \bar{p}\pi^{+}) \cdot {\cal
  B}(\omega\rightarrow \pi^{+}\pi^{-}\pi^{0})\cdot {\cal
  B}(\pi^{0}\rightarrow \gamma\gamma)$~\cite{Zyla:2020zbs}. The red
and black solid curves in Fig.~$\ref{uplimits}$ show the updated and
the raw likelihood distributions, respectively. The upper limit of the
number of signal events is 80.1, and the upper limit of the branching
fraction is $14\times 10^{-6}$ at the $90\%$ CL.

\section{Systematic Uncertainties}
\begin{table*}[phtb]
	\centering
	\caption{Systematic uncertainties in \%.}
	\begin{tabular}{llcc}\hline\hline
		\multicolumn{2}{l}{Source} & $\psi(3686)\rightarrow\Lambda\bar{\Lambda}\omega$ & Upper limit of $\Lambda^{\ast}/\bar{\Lambda}^{\ast}$\\
		\hline
		\multicolumn{2}{l}{Number of $\psi(3686)$ events}&
                $0.7$ & $0.7$\\
		\multicolumn{2}{l}{MDC tracking} & $6.0$ & $6.0$\\
		\multicolumn{2}{l}{PID efficiency}& $4.0$ & $4.0$\\
		\multicolumn{2}{l}{Photon detection efficiency} & $2.0$ & $2.0$\\
		\multicolumn{2}{l}{$\Lambda\bar{\Lambda}$ reconstruction efficiency} & $2.0$ & $2.0$\\
		\multicolumn{2}{l}{Kinematic Fit} & $1.5$ & $1.5$\\
		\cline{2-4}
		\multirow{4}{*}{Intermediate decay}
		& $\Lambda\rightarrow p\pi^{-}$ & $0.8$ & $0.8$\\
		& $\bar{\Lambda}\rightarrow \bar{p}\pi^{+}$ & $0.8$ & $0.8$\\
		& $\omega\rightarrow \pi^{+}\pi^{-}\pi^{0}$ & $0.7$ & $0.7$\\
		& $\pi^{0}\rightarrow \gamma\gamma$ & $0.03$ & $0.03$\\
	\cline{2-4}
		\multirow{4}{*}{Mass window}
		&$\Lambda$ & $0.02$ & $0.04$\\
		&$\bar{\Lambda}$ & $0.03$ & $0.02$\\
		&$J/\psi$ & $1.6$ & $1.8$\\
		&$\omega$ & - & $0.02$\\
		\cline{2-4}
		\multirow{3}{*}{Fitting}
		& Signal shape & $2.9$ & -\\
		& Background shape & $1.4$ & -\\
		& Fit range & $1.5$ & -\\
		\hline
		Total& &$8.9$&$8.2$\\
		\hline\hline
	\end{tabular}
	\label{sumsysuncertainties}
\end{table*}
The sources of systematic uncertainties are summarized
in~Table~\ref{sumsysuncertainties}. The second and third columns present the
uncertainties for the branching fraction measurement of
$\psi(3686)\rightarrow\Lambda\bar{\Lambda}\omega$ and the upper limit of ${\cal
  B}(\psi(3686)\rightarrow\Lambda\bar{\Lambda}^{\ast}+c.c~\rightarrow\Lambda\bar{\Lambda}\omega)$,
respectively.

\textit{Number of $\psi(3686)$ events}. The uncertainty due to the
number of $\psi(3686)$ events is $0.7\%$~\cite{BESIII:2017tvm}.

\textit{Tracking efficiency}. The uncertainty due to data-MC
difference in the tracking efficiency is 1.0$\%$ for each charged
track coming from a primary vertex, according to a study based on
$J/\psi \rightarrow K^{\ast}\bar{K}$ and $J/\psi \rightarrow
p\bar{p}\pi^{+}\pi^{-}$ events~\cite{BESIII:2012koo}. For tracks coming from
$\Lambda(\bar{\Lambda})$ decays, the uncertainty is also 1.0$\%$,
estimated by an analysis of $J/\psi\rightarrow \bar{p}K^{+}\Lambda$
events~\cite{BESIII:2012koo}.

\textit{PID efficiency}. In this analysis, the particle identification
is applied to the $p$($\bar{p}$) from $\Lambda(\bar{\Lambda})$ decays,
and to the $\pi^{+}$ and $\pi^{-}$ from $\omega$ decays. The PID
efficiency has been investigated using control samples of
$J/\psi\rightarrow K_{S}^{0}K^{\pm}\pi^{\pm}$ and $J/\psi\rightarrow
p\bar{p}\pi^{+}\pi^{-}$~\cite{BESIII:2011ysp,BESIII:2011wmh}. The
uncertainty is assigned to be 1.0$\%$ per charged track; thus the
total systematic uncertainty is 4.0$\%$.

\textit{Photon detection efficiency}. The uncertainty in the photon
reconstruction is studied by using the control sample
$\psi(3686)\rightarrow\pi^{+}\pi^{-}J/\psi$, $J/\psi\rightarrow
\rho^{0}\pi^{0}$, and a 1.0$\%$ systematic uncertainty is estimated
for each photon~\cite{BESIII:2011ysp}.

\textit{$\Lambda\bar{\Lambda}$ reconstruction efficiency}. The
uncertainties due to the $\Lambda$ and $\bar{\Lambda}$ secondary
vertex fits are determined to be 1.0$\%$ each, as presented in
Ref.~\cite{BESIII:2012jve}. In this work, a $2.0\%$ systematic uncertainty is estimated for the $\Lambda\bar{\Lambda}$ reconstruction efficiency.

\textit{Kinematic fit}. The systematic uncertainty due to kinematic fitting is estimated by correcting the helix parameters of charged tracks according to the method described in Ref.~\cite{BESIII:2012mpj}. The difference between the detection efficiencies with and without this correction is considered as the systematic uncertainty, which is 1.5$\%$. The detection efficiencies with this correction is nominal result. 

\textit{Intermediate decays}. The systematic uncertainties on the
intermediate-decays $\Lambda\rightarrow p\pi^{-}$,
$\bar{\Lambda}\rightarrow \bar{p}\pi^{+}$, $\omega\rightarrow
\pi^{+}\pi^{-}\pi^{0}$ and $\pi^{0}\rightarrow \gamma\gamma$ are
 from the PDG~\cite{Zyla:2020zbs}.

\textit{Mass window}. The systematic uncertainty from the requirement on the  $\Lambda$($\bar{\Lambda},\omega $) signal region is estimated by smearing the $p\pi^{-}$($\bar{p}\pi^{+}$, $\pi^{+}\pi^{-}\pi^{0}$) invariant mass in the signal MC simulation with a Gaussian function to compensate for the resolution difference between data and MC simulation. The smearing parameters are determined by fitting the $p\pi^{-}$($\bar{p}\pi^{+}$, $\pi^{+}\pi^{-}\pi^{0}$) invariant mass distribution in data with the MC-simulated shape convolved with a Gaussian function. The difference in the detection efficiency as determined from the signal MC simulation with and without the extra smearing is taken as the systematic uncertainty. The systematic uncertainties related to the vetoed $J/\psi$ mass window are estimated by varying the size of the mass window, i.e. contracting/expanding it by 2 MeV$/c^{2}$. The resulting differences of branching fractions are treated as the systematic uncertainties.

\textit{Fit range}. To estimate the systematic uncertainty due to the fit range, several alternative fits in different ranges~([0.69, 0.89]~${\rm GeV}/c^2$,~[0.71, 0.91]~${\rm GeV}/c^2$,~[0.69, 0.91]~${\rm GeV}/c^2$ and [0.71, 0.89]~${\rm GeV}/c^2$) are performed. The largest resulting difference in the BF is assigned as the systematic uncertainty.

\textit{Signal shape}. To estimate the uncertainty due to the choice of the signal shape, the MC-simulated shape is replaced by MC-simulated shape convolved with a Gaussian function, and the resulting differences in the BFs are assigned as systematic uncertainties.

\textit{Background shape}. To estimate the systematic uncertainty due to choice of the background shape, an alternative fit is performed by replacing the first order Chebyshev polynomial with a
second order Chebyshev polynomial. The change in the measured BF is assigned as the corresponding systematic uncertainty.

The total systematic uncertainty is calculated by assuming the individual components to be independent, and adding their magnitudes in quadrature.

\section{Summary}

The process $\psi(3686)\rightarrow\Lambda\bar{\Lambda}\omega$ is
observed for the first time, using $(448.1\pm2.9)\times10^6$
$\psi(3686)$ events collected with the BESIII detector. The branching fraction ${\cal
  B}(\psi(3686)\rightarrow\Lambda\bar{\Lambda}\omega)$ is measured to
be $\rm (3.30\pm0.34(stat.)\pm0.29(syst.))\times10^{-5}$. In addition,
the potential excited $\Lambda$ states are investigated by an unbinned maximum likelihood fit to the Dalitz plot. The results hint in agreement with a resonant structure with a mass around 2~${\rm GeV}/c^2$, which could be explained as an excited state $\Lambda^{*}$, but its significance (3.0$\sigma$) is not sufficient to claim an observation. The corresponding upper limit for the branching fraction ${\cal
  B}(\psi(3686)\rightarrow\Lambda\bar{\Lambda}^{\ast}+c.c~\rightarrow\Lambda\bar{\Lambda}\omega)$
is determined to be $1.40\times10^{-5}$ at the 90$\%$ confidence level.

\section{Acknowledgement}

The BESIII collaboration thanks the staff of BEPCII and the IHEP computing center for their strong support. This work is supported in part by National Key Research and Development Program of China under Contracts Nos. 2020YFA0406300, 2020YFA0406400; National Natural Science Foundation of China (NSFC) under Contracts Nos. 11625523, 11635010, 11735014, 11822506, 11835012, 11922511, 11935015, 11935018,  11961141012, 12022510, 12025502, 12035009, 12035013, 12061131003, 12147217; the Chinese Academy of Sciences (CAS) Large-Scale Scientific Facility Program; Joint Large-Scale Scientific Facility Funds of the NSFC and CAS under Contracts Nos.  U2032110, U1732263, U1832207; CAS Key Research Program of Frontier Sciences under Contract No. QYZDJ-SSW-SLH040; 100 Talents Program of CAS;
Program of Science and Technology Development Plan of Jilin Province of China under Contract No. 20210508047RQ; INPAC and Shanghai Key Laboratory for Particle Physics and Cosmology; ERC under Contract No. 758462; European Union Horizon 2020 research and innovation programme under Contract No. Marie Sklodowska-Curie grant agreement No 894790; German Research Foundation DFG under Contracts Nos. 443159800, Collaborative Research Center CRC 1044, FOR 2359, GRK 2149; Istituto Nazionale di Fisica Nucleare, Italy; Ministry of Development of Turkey under Contract No. DPT2006K-120470; National Science and Technology fund; Olle Engkvist Foundation under Contract No. 200-0605; STFC (United Kingdom); The Knut and Alice Wallenberg Foundation (Sweden) under Contract No. 2016.0157; The Royal Society, UK under Contracts Nos. DH140054, DH160214; The Swedish Research Council; U. S. Department of Energy under Contracts Nos. DE-FG02-05ER41374, DE-SC-0012069.


\begin{thebibliography}{99}

\bibitem{BESIII:2018gmc}
M.~Ablikim \textit{et al.} [BESIII],
\href{https://doi.org/10.1103/PhysRevD.99.032006}{Phys. Rev. D \textbf{99}, 032006 (2019).}

\bibitem{Zhu:2015bha}
K.~Zhu, X.~H.~Mo and C.~Z.~Yuan,
\href{https://doi.org/10.1142/S0217751X15501481}{Int. J. Mod. Phys. A \textbf{30}, 1550148 (2015).}

\bibitem{BESIII:2012jve}
M.~Ablikim \textit{et al.} [BESIII Collaboration],
\href{https://doi.org/10.1103/PhysRevD.87.052007}{Phys. Rev. D \textbf{87}, 052007 (2013).}

\bibitem{Sarantsev:2019xxm}
A.~V.~Sarantsev, M.~Matveev, V.~A.~Nikonov, A.~V.~Anisovich, U.~Thoma and E.~Klempt,
\href{https://doi.org/10.1140/epja/i2019-12880-5}{Eur. Phys. J. A \textbf{55}, 180 (2019).}

\bibitem{Zyla:2020zbs}
P.~A.~Zyla \textit{et al.} [Particle Data Group],
\href{https://doi.org/10.1093/ptep/ptaa104}{Prog. Theor. Exp. Phys.~\textbf{2020}, 083C01 (2020).}

\bibitem{BESIII:2017tvm}
M.~Ablikim \textit{et al.} [BESIII Collaboration],
\href{https://doi.org/10.1088/1674-1137/42/2/023001}{Chin. Phys. C \textbf{42}, 023001 (2018).}

\bibitem{Ablikim:2009aa}
M.~Ablikim {\it et al.} [BESIII Collaboration],
\href{https://doi.org/10.1016/j.nima.2009.12.050}{Nucl. Instrum.
Methods Phys. Res., Sect. A {\bf 614}, 345 (2010).}

\bibitem{Yu:IPAC2016-TUYA01}
   C.~H.~Yu {\it et al.},
  Proceedings of IPAC2016, Busan, Korea (JACoW, Busan, 2016),
  \url{https://accelconf.web.cern.ch/ipac2016/}.


\bibitem{BESIII:2020nme}
M.~Ablikim \textit{et al.} [BESIII Collaboration],
\href{https://doi.org/10.1088/1674-1137/44/4/040001}{Chin. Phys. C \textbf{44},  040001 (2020).}

\bibitem{GEANT4:2002zbu}
S.~Agostinelli \textit{et al.} [GEANT4 Collaboration],
\href{https://doi.org/10.1016/S0168-9002(03)01368-8}{Nucl. Instrum. Meth. A \textbf{506}, 250 (2003).}

\bibitem{Jadach:2000ir}
S.~Jadach, B.~F.~L.~Ward and Z.~Was,
\href{https://doi.org/10.1103/PhysRevD.63.113009}{Phys. Rev. D \textbf{63}, 113009 (2001).}

\bibitem{Jadach:1999vf}
S.~Jadach, B.~F.~L.~Ward and Z.~Was,
\href{https://doi.org/10.1016/S0010-4655(00)00048-5}{Comput. Phys. Commun. \textbf{130}, 260 (2000).}

\bibitem{Lange:2001uf}
D.~J.~Lange,
\href{https://doi.org/10.1016/S0168-9002(01)00089-4}{Nucl. Instrum. Meth. A \textbf{462}, 152 (2001).}


\bibitem{Ping:2008zz}
R.~G.~Ping,
\href{https://doi.org/10.1088/1674-1137/32/8/001}{Chin. Phys. C \textbf{32}, 599 (2008).}

\bibitem{Chen:2000tv}
J.~C.~Chen, G.~S.~Huang, X.~R.~Qi, D.~H.~Zhang and Y.~S.~Zhu,
\href{https://doi.org/10.1103/PhysRevD.62.034003}{Phys. Rev. D \textbf{62}, 034003 (2000).}

\bibitem{Yang:2014vra}
R.~L.~Yang, R.~G.~Ping and H.~Chen,
\href{https://doi.org/10.1088/0256-307X/31/6/061301}{Chin. Phys. Lett. \textbf{31}, 061301 (2014).}

\bibitem{Richter-Was:1992hxq}
E.~Richter-Was,
\href{https://doi.org/10.1016/0370-2693(93)90062-M}{Phys. Lett. B \textbf{303}, 163 (1993).}

\bibitem{Ablikim:2013ntc}
M.~Ablikim [BESIII Collaboration],
\href{https://doi.org/10.1088/1674-1137/37/12/123001}{Chin. Phys. C \textbf{37}, 123001 (2013).}

\bibitem{BESIII:2019efv}
M.~Ablikim \textit{et al.} [BESIII Collaboration],
\href{https://doi.org/10.1103/PhysRevD.100.052010}{Phys. Rev. D \textbf{100}, 052010 (2019).}




\bibitem{BESIII:2017qwj}
M.~Ablikim \textit{et al.} [BESIII Collaboration],
\href{https://doi.org/10.1016/j.physletb.2017.05.033}{Phys. Lett. B \textbf{771}, 45 (2017).}

\bibitem{Zhou:2020ksj}
X.~Zhou, S.~Du, G.~Li and C.~Shen,
\href{https://doi.org/10.1016/j.cpc.2020.107540}{Comput. Phys. Commun. \textbf{258}, 107540 (2021).}

\bibitem{BESIII:2019efv}
M.~Ablikim \textit{et al.} [BESIII Collaboration],
\href{https://doi.org/10.1103/PhysRevD.100.052010}{Phys. Rev. D \textbf{100}, 052010 (2019).}

\bibitem{BESIII:2017vtc}
M.~Ablikim \textit{et al.} [BESIII Collaboration],
\href{https://doi.org/10.1103/PhysRevD.97.052001}{Phys. Rev. D \textbf{97}, 052001 (2018).}

\bibitem{BESIII:2018gvg}
M.~Ablikim \textit{et al.} [BESIII Collaboration],
\href{https://doi.org/10.1103/PhysRevD.98.032014}{Phys. Rev. D \textbf{98}, 032014 (2018).}

\bibitem{BESIII:2017tvm}
M.~Ablikim \textit{et al.} [BESIII Collaboration],
\href{https://doi.org/10.1088/1674-1137/42/2/023001}{Chin. Phys. C \textbf{42}, 023001 (2018).}

\bibitem{BESIII:2012koo}
M.~Ablikim \textit{et al.} [BESIII Collaboration],
\href{https://doi.org/10.1103/PhysRevD.87.012007}{Phys. Rev. D \textbf{87}, 012007 (2013).}

\bibitem{BESIII:2011ysp}
M.~Ablikim \textit{et al.} [BESIII Collaboration],
\href{https://doi.org/10.1103/PhysRevD.83.112005}{Phys. Rev. D \textbf{83}, 112005 (2011).}

\bibitem{BESIII:2011wmh}
M.~Ablikim \textit{et al.} [BESIII Collaboration],
\href{https://doi.org/10.1103/PhysRevD.85.092012}{Phys. Rev. D \textbf{85}, 092012 (2012).}


\bibitem{BESIII:2012mpj}
M.~Ablikim \textit{et al.} [BESIII Collaboration],
\href{https://doi.org/10.1103/PhysRevD.87.012002}{Phys. Rev. D \textbf{87}, 012002 (2013).}
\end{thebibliography}
\end{document}